\documentclass[aps,prb,reprint,longbibliography]{revtex4-1}
\usepackage[utf8]{inputenc}
\usepackage{hyperref}
\usepackage{amsmath}
\usepackage{graphicx}

\newcommand{\sci}[2]{\ensuremath{#1 \times 10^{#2}}}
\newcommand{\rmd}{\ensuremath{\mathrm{d}}}
\newcommand{\unit}{1\!\!\!\!\;1}

\begin{document}
\title{Electron mobility in graphene without invoking the Dirac equation}
\author{Chaitanya K. Ullal}
\author{Jian Shi}
\author{Ravishankar Sundararaman}\email{sundar@rpi.edu}
\affiliation{Department of Materials Science and Engineering, Rensselaer Polytechnic Institute, 110 8$^{th}$ Street, Troy, NY 12180}

\begin{abstract}
The Dirac point and linear band structure in Graphene bestow it with remarkable
electronic and optical properties, a subject of intense ongoing research.
Explanations of high electronic mobility in graphene, often invoke the
masslessness of electrons based on the effective relativistic Dirac-equation behavior,
which are inaccessible to most undergraduate students and are not intuitive
for non-physics researchers unfamiliar with relativity.
Here, we show how to use only basic concepts from semiconductor theory
and the linear band structure of graphene to explain its unusual
effective mass and mobility, and compare them with conventional
metals and semiconductors.
We discuss the more intuitive concept of transverse effective mass
that emerges naturally from these basic derivations,
which approaches zero in the limit of undoped graphene at low temperature
and is responsible for its extremely high mobility.
\end{abstract}

\maketitle

\section{Introduction}
Graphene is often described in superlatives, with a multitude of extreme
electronic, mechanical and chemical properties of interest in disparate
fields of research.\cite{GrapheneRise,GrapheneConclusions,CarbonWonderland}
This increasingly motivates exposure to graphene science at the undergraduate level,\cite{AJPNanotechResourceLetter}
with excellent pedagogical resources introducing the calculation of its unique
Dirac-point band structure,\cite{BandStructMatrixMechanics,ChainModel1D}
explaining novel transport phenomena such as Klein tunneling,\cite{KleinTunneling}
and even outlining experimental demonstrations of the unique wave mechanics
of honeycomb lattices in ripple tanks.\cite{WavePhysicsRippleTank}

\begin{figure}
\includegraphics[width=\columnwidth]{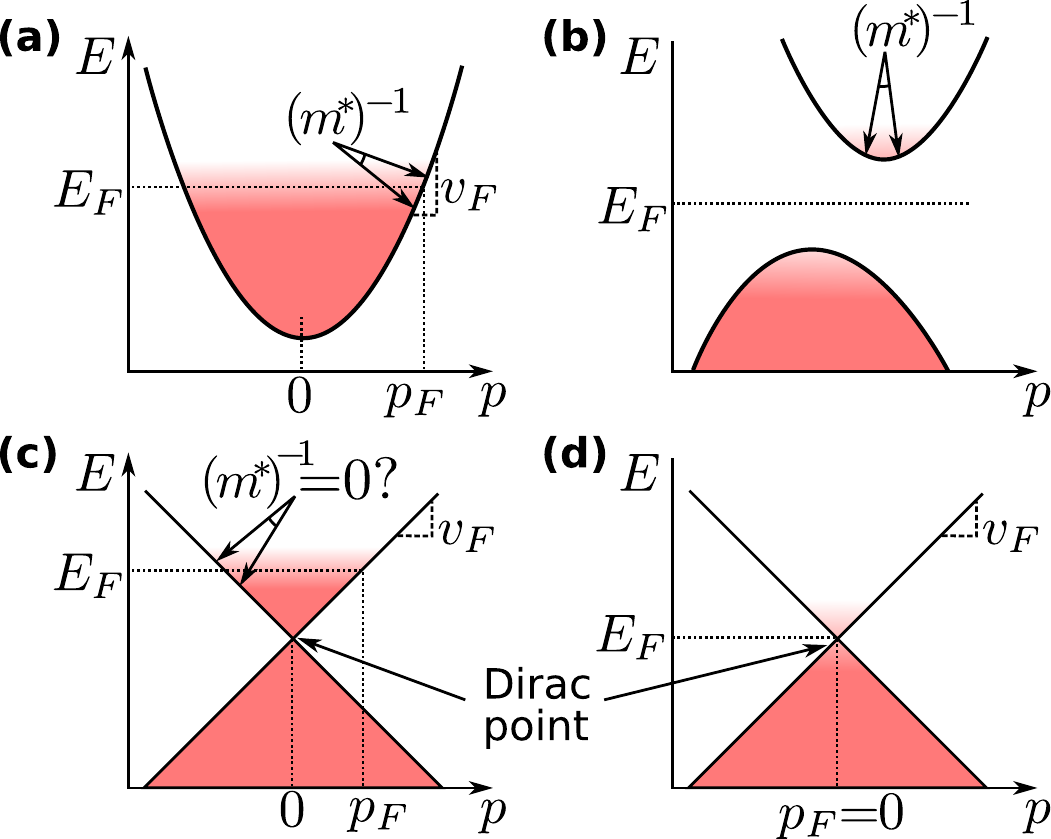}
\caption{Schematic band structure of (a) a free -electron metal,
(b) a parabolic-band semiconductor, (c) doped graphene
and (d) undoped graphene, with band velocity ($E$ vs $p$ slopes)
and effective mass (inverse $E$ vs $p$ curvature) annotated.
The shading denotes the Fermi occupation factors of electrons.
Naively, the linear band structure yields zero curvature and an
infinite effective mass in graphene, rather than zero or a low value.
\label{fig:BandStruct}}
\end{figure}

Of graphene's extreme properties, its exceptional electrical conductivity and mobility,
arising from the effective masslessness of electrons in the Dirac band structure
are often discussed.\cite{MasslessElectrons}
Explaining the high mobility from a low effective mass is easily accessible
at an undergraduate level with standard semiconductor physics derivations
of Drude theory.\cite{SemiconductorPhySze} However, explaining why the Dirac
band structure corresponds to massless carriers is somewhat more challenging,
and has not yet been discussed clearly in a pedagogical context.
Specifically, the Dirac band structure contains a linear dispersion
relation $E = v_F p$, corresponding to a constant electron speed
$v = \partial E/\partial p = v_F$, independent of the momentum.
Here, $v_F\approx \sci{8.3}{5}$~m/s is the Fermi velocity of graphene,
the velocity of electrons at the Fermi energy up to which states
are filled with electrons.\cite{AshcroftMermin}
Naive application of the conventional semiconductor definition of effective mass,
which corresponds to $(m^\ast)^{-1} = \partial v/\partial p = 0$ as we discuss below
in further detail, leads to the opposite result of infinite mass!
(See Fig.~\ref{fig:BandStruct}.)

Research papers invoke seminal work\cite{NovoselovNature2005} that demonstrated
that electron transport in graphene is essentially governed by the Dirac equation,
with the charge carriers mimicking relativistic particles with zero rest mass.
In the relativistic picture, linear dispersion corresponds to
massless carriers by recognizing that, $E^2 = (mc^2)^2 + (pc)^2$
for a particle of mass $m$, reduces to $E = pc$ when $m=0$.
However, this is only an analogy\cite{GrapheneReview}
and must be applied very carefully to graphene.
For graphene, the electrons have a constant velocity $v_F 
\approx \sci{8.3}{5}$~m/s $\sim c/400$, as discussed above,
instead of the speed of light $c$.
Importantly, there is no Lorentz invariance for carriers in graphene:
the frame in which the carbon nuclei are at rest is special!
Therefore, explaining masslessness of graphene carriers using relativity,
though valid when done correctly, will likely lead to confusion
especially at the introductory undergraduate level.
Moreover, it is not an intuitive explanation for students from related
fields in chemistry, materials science or electrical engineering,
who are all increasingly likely to encounter graphene in their careers.

Pedagogical descriptions that try to avoid the relativistic / Dirac explanation
often rely on alternate definitions of the mass that work correctly
for graphene e.g. cyclotron effective mass\cite{CyclotronMass,SolidStateHoffman},
quarternion effective mass\cite{Quaternions} etc.
While these definitions work for a reason, as we will discuss below,
they do not provide an intuitive picture of how electrons in graphene conduct remarkably well.
Most importantly, some of these alternate approaches attempt to redefine
the effective mass as the ratio of the \emph{momentum to velocity},\cite{CyclotronMass}
$m^\ast = p/v$, rather than the ratio of \emph{force to acceleration},
$m^\ast = F/a = \partial p/\partial v$. Such alternate definitions are
correct only for a parabolic band structure, where the mass is independent
of the momentum and the two expressions above become equivalent.

Here we describe an alternate pedagogical approach to explain massless electrons in graphene,
where we retain the standard definition of effective mass from semiconductor theory, albeit the full tensorial version.
We demonstrate how to work through this definition to understand how graphene's effective mass and
mobility vary with doping and temperature, and how it contrasts with conventional metals and semiconductors.
We show how to arrive at the concept that the transverse effective mass,
rather than the usual longitudinal one, dominates transport in graphene,
and is the mass that approaches zero near the Dirac point in graphene.
The approach presented here should be suitable for intuitively explaining
the remarkable electronic properties of graphene, a topic of continuing
research interest, at the senior undergraduate level.

\section{Derivations}

Electrons in materials with a band structure or dispersion relation $E(k)$
have group velocity $v = \partial E/\partial p$ and momentum $p = \hbar k$,
using only basic principles in quantum mechanics.
When an electric field is applied to the material, the electric
force accelerates the electrons and generates a current.
For the same force, lighter electrons will be accelerated more,
and will result in higher mobility and conductivity.
Consequently, the mass relevant for determining conduction by electrons is
the ratio of force $F$ to acceleration $a$ (exactly as in Newton's second law).
Now $F = \rmd p/\rmd t$ and $a = \rmd v/\rmd t$, which yield
\begin{flalign}
(m^\ast)^{-1} 
&\equiv \frac{\rmd v/\rmd t}{\rmd p/\rmd t}
= \frac{\partial v}{\partial p}
= \frac{\partial (\partial E/\partial p)}{\partial p}
\nonumber\\
&\equiv \frac{\partial^2 E}{\partial p^2}
= \frac{\partial^2 E}{\hbar^2 \partial k^2},
\end{flalign}
the well-known expression in semiconductor theory that the effective mass is the inverse
of the curvature of the band structure $E(k)$.\cite{AshcroftMermin,SemiconductorPhySze}
The curvature and effective mass are both finite (non-zero and not infinite)
for metals and semiconductors, as shown in Fig.~\ref{fig:BandStruct}.
However, for graphene, the linear band structure has seemingly zero curvature
corresponding to an infinite effective mass, in stark contrast to the
massless carrier explanation for its high mobility.

The simplest correct explanation for massless electrons in graphene
lies within the standard definition, but necessitates the full
tensorial version,\cite{AshcroftMermin,SemiconductorPhySze}
\begin{equation}
(\bar{m}^\ast)^{-1} \equiv \frac{1}{\hbar^2} \nabla_{\vec{k}} \nabla_{\vec{k}} E(\vec{k}).
\end{equation}
The mass tensor is just a matrix that connects how changes of momentum
and velocity are related, $\rmd\vec{p} = \bar{m}^\ast \cdot \rmd\vec{v}$,
or equivalently $\rmd\vec{v} = (\bar{m}^\ast)^{-1} \cdot \rmd\vec{p}$,
which are not in the same direction for a general $E(\vec{k})$.
For two-dimensional graphene near the Dirac point,
the above definition reduces to
\begin{equation}
(\bar{m}^\ast)^{-1} = \frac{1}{\hbar^2}
\left(\begin{array}{cc}
\partial_{k_x}^2 & \partial_{k_x}\partial_{k_y} \\
\partial_{k_x}\partial_{k_y} & \partial_{k_y}^2
\end{array}\right)
 \hbar v_F \sqrt{k_x^2 + k_y^2}.
\end{equation}
Straightforward evaluation of the derivatives yields
\begin{align}
(\bar{m}^\ast)^{-1}
&= \frac{v_F}{\hbar}
	\left(\begin{array}{cc}
	\frac{k_y^2}{(k_x^2 + k_y^2)^{3/2}} & -\frac{k_x k_y}{(k_x^2 + k_y^2)^{3/2}} \\
	-\frac{k_x k_y}{(k_x^2 + k_y^2)^{3/2}} & \frac{k_x^2}{(k_x^2 + k_y^2)^{3/2}}
	\end{array}\right) \\
&= \frac{v_F}{p^3}
\left(\begin{array}{cc}
	p_y^2 & -p_x p_y \\
	-p_x p_y & p_x^2
	\end{array}\right),
\end{align}
using $\vec{p} \equiv \hbar\vec{k}$.
With the definition
\begin{equation}
\bar{M}(\phi) \equiv
\left(\begin{array}{cc}
\sin^2\phi & -\sin\phi\cos\phi \\
-\sin\phi\cos\phi & \cos^2\phi
\end{array}\right), \label{eqn:Mphi}
\end{equation}
the inverse effective mass tensor can be written in polar coordinates as
\begin{equation}
(\bar{m}^\ast)^{-1} = \frac{v_F}{p} \bar{M}(\phi). \label{eqn:mAstGeneral}
\end{equation}

As with any symmetric tensor, the inverse mass tensor is best characterized
in its principal axes or eigenvectors, so that it becomes diagonal.
Solving the characteristic equation $\det \left[ (\bar{m}^\ast)^{-1} - \lambda \unit \right] = 0$
yields the two eigenvalues
\begin{equation}
\lambda = \left\{ 0, \frac{v_F}{p} \right\},
\end{equation}
and their corresponding eigenvectors can be derived to be
\begin{equation}
\vec{x} = \left\{
	\frac{1}{p} \left(\begin{array}{c} p_x \\ p_y \end{array} \right),
	\frac{1}{p} \left(\begin{array}{c} p_y \\ -p_x \end{array} \right)
\right\}.
\end{equation}
The first eigenvector is exactly $\hat{p}$, the unit vector along the momentum.
This principal direction therefore corresponds to changes in momentum
parallel to the momentum direction, which is a `longitudinal' change.
The corresponding inverse mass eigenvalue is 0, therefore implying
that the longitudinal mass $m_L \to \infty$, which is exactly the
result we obtained in the non-tensorial analysis.

However, now we have the second eigenvector which is perpendicular to $\hat{p}$,
corresponding to changes in momentum perpendicular to the momentum direction,
which is a `transverse' change.
The corresponding inverse mass eigenvalue is $v_F/p$,
corresponding to a transverse mass $m_T = p/v_F$.
As we approach the Dirac point $p \to 0$,
the transverse mass $m_T \to 0$.
Therefore, at the Dirac point, electrons in graphene have an infinite
longitudinal mass, but are massless in the transverse direction.

\begin{figure}
\includegraphics[width=\columnwidth]{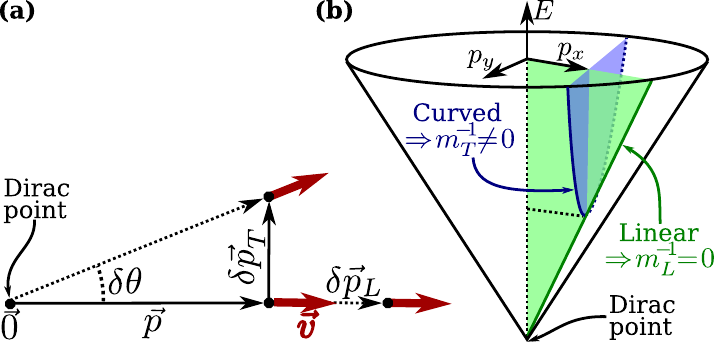}
\caption{(a) Velocity (thick red arrows) is always parallel
to momentum (thin black arrows) with constant magnitude $v_F$.
Therefore, velocity is unchanged for longitudinal changes in
momentum $\delta\vec{p}_L$ yielding infinite longitudinal mass $m_L$.
Velocity changes direction for transverse changes $\delta\vec{p}_T$,
resulting in a small transverse mass $m_T$ which $\to 0$ as $p \to 0$.
(b) Linear $E(k)$ in the radial (longitudinal) slice of the
conical $E(\vec{k})$ yields $m_L \to \infty$, while parabolic $E(k)$
in the transverse slice yields a small $m_T$ that approaches zero
as the slice gets closer to the Dirac point.
\label{fig:VectorMath}}
\end{figure}

This result can also be understood intuitively directly from the
linear dispersion relation, $E(\vec{p}) = v_F |\vec{p}|$.
The corresponding velocity $\vec{v} \equiv \nabla_{\vec{p}} E(\vec{p}) = v_F \hat{p}$,
which is always parallel to the momentum $\vec{p}$,
but has a constant magnitude $v_F$ independent of $p$.
As shown in Fig.~\ref{fig:VectorMath}(a), when the momentum is
changed in the longitudinal direction, the momentum direction $\hat{p}$
is unchanged, so the velocity direction and magnitude remain unchanged.
No change in velocity with changing momentum yields zero inverse mass
and an infinite longitudinal mass, $m_L \to \infty$.
However, when momentum is changed in the transverse direction
by $\delta\vec{p}_T$, the momentum and velocity directions both
change by angle $\delta\theta = \delta p_T/p$.
Since the velocity magnitude is unchanged, the velocity vector
changes by $\delta v = v_F \delta\theta = v_F \delta p_T/p$.
Therefore, the ratio of velocity to momentum change is $v_F/p$
which corresponds to the transverse mass, $m_T = p/v_F$.
This can also be seen comparing radial (longitudinal) and transverse slices
of the conical $E(\vec{k})$ near the Dirac point (Fig.~\ref{fig:VectorMath}(b)).
The linear $E(k)$ along the longitudinal slice yields $m_L \to \infty$,
while the parabolic $E(k)$ in the transverse slice yields finite $m_T$.
The transverse curvature increases as the slice gets closer to the
Dirac point, resulting in $m_T \to 0$ near the Dirac point.

The transverse mass is closely related to the cyclotron mass,
which is the reason why the latter definition works for graphene.\cite{CyclotronMass}
In a magnetic field, charged particles move in circles with
centripetal force and acceleration perpendicular to the velocity.
The cyclotron mass is the ratio of force to acceleration when they
are both perpendicular to the momentum (velocity) direction,
which is exactly the case for the transverse mass as discussed above.

When an electric field $\vec{E}$ is applied to a material,
this applies a force $-e\vec{E}$ on all the electrons,
resulting in an acceleration $\vec{a} = -e(\bar{m}^\ast)^{-1}\cdot\vec{E}$.
As the electrons move through the material, they scatter
against defects and lattice vibrations (phonons)
which causes the velocity to randomize due to collisions
over the Drude relaxation time scale $\tau$.
With these two effects, the electrons pick up an
average drift velocity $\vec{v}_d = \vec{a}\tau
= -e\tau(\bar{m}^\ast)^{-1}\cdot\vec{E}$.
The ratio of drift velocity to the applied electric field
defines the mobility, (excluding the sign due to negative charge)
\begin{equation}
\bar{\mu} = e\tau(\bar{m}^\ast)^{-1}.
\end{equation}
The total current density in the electron is $\vec{j} = n(-e)\vec{v}_d
= ne\bar{\mu}\cdot\vec{E}$, where $n$ is the number density of electrons.
The conductivity (tensor) is defined by $\vec{j} = \bar{\sigma}\cdot\vec{E}$,
which implies $\bar{\sigma} = ne\bar{\mu}$, elucidating
the mobility to be the conductivity per unit charge density.

In general, the effective mass $m^\ast$ varies with $\vec{k}$
(i.e. $\vec{p}$), and hence so does the mobility.
The experimentally-determined mobility is therefore
an average over all charge carriers.
First, consider the case of n-doped graphene where a net excess
of electrons over holes results in states being occupied
up to an energy $E_F$ above the Dirac point.
(The discussion for excess holes with $E_F$ below the Dirac point
follows in exactly the same way, with exactly the same results,
due to the electron $\leftrightarrow$ hole symmetry in the band structure.)
From the linear energy relation $E = v_F p$, we can see that this
corresponds to a Fermi momentum $p_F = E_F/v_F$ and
wave-vector $k_F = E_F/(\hbar v_F)$. (Fig.~\ref{fig:BandStruct}(c))

\begin{figure}
\includegraphics[width=0.7\columnwidth]{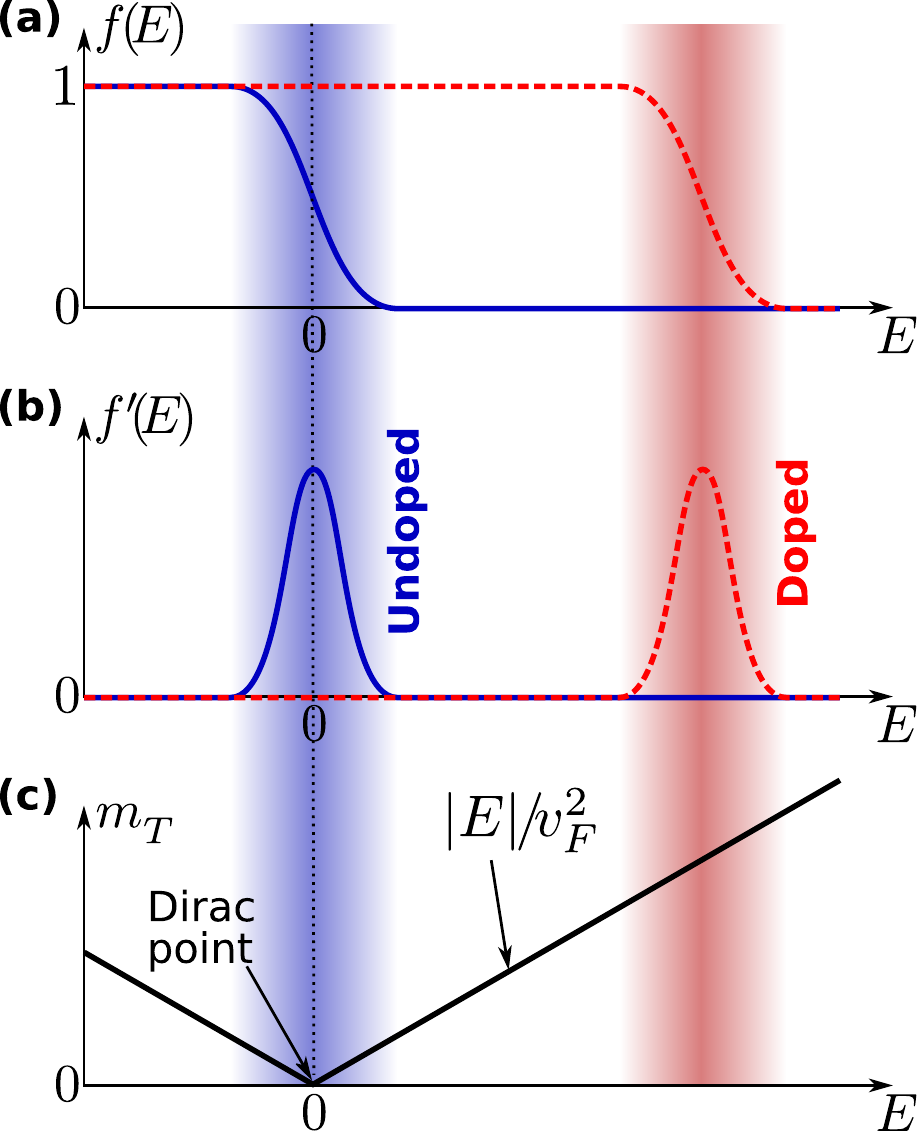}
\caption{(a) Fermi function and (b) its derivative for
undoped and doped graphene. Shaded regions indicate the contribution
to conduction, proportional to the Fermi function derivative.
(c) Transverse effective mass in graphene as a function of
electron energy, which is distributed around a non-zero value
for doped graphene, but around zero for undoped graphene.
\label{fig:FermiFunctions}}
\end{figure}

Only electrons within a few $k_BT$ of the Fermi energy contribute
to electronic conduction in materials. Intuitively, only these
electrons have empty states available to `move' to, 
being close to the energy at which electronic states
transition from filled to empty (Fig.~\ref{fig:FermiFunctions}(a)).
In fact, a more detailed analysis based on the Boltzmann transport
equation and the relaxation time approximation\cite{AshcroftMermin}
shows that the contribution to conduction is proportional to the
derivative of the Fermi function (Fig.~\ref{fig:FermiFunctions}(b)).
If the number density of electrons due to doping is high enough
that $E_F \gg k_BT$, then all the electrons that contribute to
conduction have approximately the same magnitude of momentum, $p_F$.
Correspondingly, they all have transverse mass $m_T \approx
p_F/v_F = E_F/v_F^2$ (Fig.~\ref{fig:FermiFunctions}(c)),
while the longitudinal mass remains $\infty$ (as shown above for all graphene electrons).

When the effective mass is anisotropic, the well-known simplified
expression for mobility $\mu = e\tau/m^\ast$ remains valid
provided an appropriate average of $m^\ast$ in all directions is used.
In particular, $m^\ast$ should be the harmonic mean of all directions
i.e. if the contributions are $m_1$, $m_2$ and $m_3$ in three
perpendicular directions (principal axes), the net effective mass
should be $(m^\ast)^{-1} = (m_1^{-1} + m_2^{-1} + m_3^{-1})/3$.
For example, in Silicon, $m_L = 0.89$ and there are two equal
transverse values (in 3D) $m_{T1} = m_{T2} = 0.19$.
The corresponding average value for mobility will then be $m^\ast = 0.26$.
For graphene, we now have $m_L = \infty$ and $m_T = p_F/v_F$ (just one in 2D),
which yields 
\begin{equation}
m^\ast = \frac{2}{m_L^{-1} + m_T^{-1}} = \frac{2p_F}{v_F} = \frac{2E_F}{v_F^2},
\end{equation}
which is twice the transverse value.
Correspondingly, we expect the mobility to be $e\tau/m^\ast = e\tau v_F^2/(2E_F)$.

We can alternatively derive this result by averaging
the mobility contributions due to all electrons
contributing to conduction on this `Fermi circle' of radius $p_F$.
This amounts to an average over $\phi$:
\begin{align}
\bar{\mu} &\equiv e\tau \frac{\int \rmd\phi (\bar{m}^\ast)^{-1}(\vec{p}_F)}{\int \rmd \phi} \\
&= e\tau \frac{\int \rmd\phi \frac{v_F}{p_F} \bar{M}(\phi)}{2\pi} \nonumber\\
&= \frac{e\tau v_F^2}{E_F} \cdot \frac{1}{2\pi} \int_0^{2\pi}\rmd\phi \bar{M}(\phi)\nonumber
&= \frac{e\tau v_F^2}{2E_F}\unit,
\end{align}
by substituting (\ref{eqn:mAstGeneral}) and (\ref{eqn:Mphi}),
and noting that the angular integrals $\int_0^{2\pi}\rmd\phi \cos^2\phi
= \int_0^{2\pi}\rmd\phi \sin^2\phi = \pi$ and $\int_0^{2\pi}\rmd\phi \cos\phi\sin\phi = 0$.
This also corresponds to an isotropic mobility $e\tau/m^\ast$,
with the effective $m^\ast = 2p_F/v_F = 2E_F/v_F^2$ as argued above.

In pure (undoped) graphene, the electronic states
switch from being occupied to unoccupied at the Dirac point
(Figs.~\ref{fig:BandStruct}(d) and \ref{fig:FermiFunctions}(a).)
The contribution to conduction, proportional to the Fermi function
derivative (Fig.~\ref{fig:FermiFunctions}(b)) as discussed above,
is centered near $E=0$.
Unlike the doped case, the corresponding effective mass is no
longer of similar magnitude throughout the energy range with
contributions to conduction, as shown in Fig.~\ref{fig:FermiFunctions}(c).
In particular, the effective mass in the center of the distribution
at the Dirac point is zero, while it is non-zero and linearly
increasing away from the Dirac point. Therefore, we need to
average over the carriers proportional to the Fermi function
derivative to estimate the mobility for undoped graphene.

For undoped graphene with the Fermi energy at the Dirac point, $E_F=0$,
the occupation of electrons is given by the Fermi function
\begin{equation}
f(E) = \frac{1}{1+\exp\frac{E}{k_BT}},
\end{equation}
with derivative
\begin{equation}
f'(E) = \frac{\exp\frac{E}{k_BT}}{k_BT\left(1+\exp\frac{E}{k_BT}\right)^2}
= \frac{1}{4k_BT}\mathrm{sech}^2\frac{E}{2k_BT}.
\end{equation}

We can therefore determine the average mobility of undoped graphene as
\begin{align}
\bar{\mu} &\equiv e\tau \frac
	{\int \rmd p_x \rmd p_y f'(v_Fp) (\bar{m}^\ast)^{-1}(\vec{p})}
	{\int \rmd p_x \rmd p_y f'(v_Fp)} \nonumber\\
&= e\tau \frac
	{\int_0^\infty p\rmd p \int_0^{2\pi}\rmd\phi~\mathrm{sech}^2\left(\frac{v_Fp}{2k_BT}\right) \frac{v_F}{p} \bar{M}(\phi)}
	{\int_0^\infty p\rmd p \int_0^{2\pi}\rmd\phi~\mathrm{sech}^2\left(\frac{v_Fp}{2k_BT}\right)} \nonumber\\
&= \frac{e\tau v_F^2}{2k_BT}
	\cdot \frac{\int_0^\infty\rmd x~\mathrm{sech}^2 x}{\int_0^\infty\rmd x~x~\mathrm{sech}^2 x}
	\cdot \frac { \int_0^{2\pi}\rmd\phi \bar{M}(\phi)}{2\pi},
\end{align}
switching the integrals over momenta to polar coordinates and
substituting $x = v_Fp/(2k_BT)$ to simplify the integral over $p$.
The integrals over $x$ are standard definite integrals that evaluate to the
constants $1$ and $\ln 2$ in the numerator and denominator respectively,
while the final term is exactly what we evaluated above to be ${\unit}/2$.
Putting that all together yields
\begin{equation}
\bar{\mu}
= \frac{e\tau v_F^2}{2k_BT} \cdot \frac{1}{\ln 2} \cdot \frac{\unit}{2}
= \frac{e\tau v_F^2}{(4\ln 2)k_BT}\unit.
\end{equation}
Note that the average mobility is isotropic (scalar) as expected
and corresponds to an averaged effective mass
\begin{equation}
\overline{m^\ast} = \frac{(4\ln 2)k_BT}{v_F^2},
\end{equation}
which is directly proportional to temperature.
This is because the transverse mass $m_T \propto p$,
and the average magnitude of momentum for electrons
in graphene at finite temperature $\propto T$.
This is in sharp contrast to conventional metals and semiconductors,
and even doped graphene with $E_F\gg k_BT$ as considered above,
where the effective mass depends only weakly on temperature.

\section{Results and Discussion}

Table~\ref{tab:MuCompare} compares the typical effective masses $m^\ast$,
momentum relaxation time $\tau$ and mobility $\mu$ of electrons
in a prototypical metal (silver), semiconductor (silicon)
and undoped graphene.
The values for silver and silicon are based on experimental measurements,
while that for graphene is based on the above derivation along with
a first-principles calculated value\cite{GraphiteHotCarriers} of $\tau \approx 2$~ps
for ideal undoped graphene and $\tau \approx 700$~fs for graphene (ideally) doped
to a Fermi energy of 0.1~eV (limited only by electron-phonon scattering).

\begin{table}
\caption{Comparison of typical electron mobility, effective mass and relaxation time
between metals, semiconductors, doped graphene and (undoped) graphene at room temperature.
\label{tab:MuCompare}}
\begin{tabular}{cccc}
\hline\hline
Material & $\tau$ [fs] & $m^\ast/m_e$ & $\mu$ [cm$^2$/V$\cdot$s]\\
\hline
Silver   &  30  &  1.0  &  50        \\
Silicon  & 200  &  0.26 & 1400       \\
Graphene with $E_F=0.1$~eV &  700 & 0.063 & \sci{2}{4} \\
Undoped Graphene           & 2000 & 0.018 & \sci{2}{5} \\
\hline\hline
\end{tabular}
\end{table}

Metals have a short relaxation time because they have a large number
of states at the Fermi level which enhances electron-phonon scattering.
Semiconductors and graphene have much smaller density of states at the
energies of electrons that carry current, resulting in an increased
relaxation time by one and two orders of magnitude relative to the metal.
The typical effective mass is somewhat smaller in semiconductors than metals,
but it is two orders of magnitude smaller at room temperature in graphene
because the transverse mass approaches zero near the Dirac point.
Consequently the mobility $\propto \tau/m^\ast$ is smallest for metals.
Semiconductor mobilities are one-two orders larger due both
to larger $\tau$ and somewhat smaller $m^\ast$.
However, in graphene both factors contribute two orders
making the mobility at room temperature four orders larger!

Note that despite the much higher mobilities in semiconductors and graphene,
the number density $n$ of electrons in metals is sufficiently larger
that the conductivity $\sigma = ne\mu$ is still much larger in metals.
Specifically, in graphene, the mobility is higher for undoped graphene
due to the lower effective mass (and additionally because of a lowered
electron-phonon scattering rate\cite{GraphiteHotCarriers}) than the doped case.
However, mobility is effectively the conductivity per carrier available for
conduction, and the number of carriers is much smaller for undoped graphene.
Consequently, undoped graphene has a low conductivity despite the highest mobility,
and graphene actually achieves a higher conductivity at an optimal doping level
where the increasing effective mass and scattering rate are compensated
by an increased carrier density.\cite{GrapheneRise,GrapheneConclusions}

As temperature changes, the scattering time $\tau$ is roughly inversely
proportional to temperature near room temperature for pure materials
because the amplitude of lattice vibrations increases with temperature.
The effective mass is mostly temperature dependent in metals and semiconductors,
so that the temperature dependence of mobility follows the scattering time.
However for undoped graphene, the effective mass is also temperature dependent
causing an additional decrease of mobility with increasing temperature
and resulting in an overall $T^{-2}$ dependence near room temperature.
The temperature dependence for doped graphene will be similar to
conventional metals and semiconductors because the average momentum
and hence the average transverse mass is set by the doping level
and not by temperature, as derived above.

\section{Conclusions}

We have presented a simplified approach to explain the remarkable mobility of graphene
that relies only on the standard semiconductor theory definition, completely avoiding
the conventionally-invoked parallel to the Dirac equation and a relativistic picture.
We discussed the calculation of the tensorial effective mass, the emergence of a zero
transverse mass (but infinite longitudinal mass) upon approaching the Dirac point
and the corresponding temperature-dependent mobilities.
The full derivations require only basic concepts from calculus, thermodynamics
and semiconductor theory, accessible to undergraduate students in
physics, chemistry, materials science and electrical engineering.
In addition, we pictorially discussed the concept of transverse effective mass
and contrast it with the more intuitively-familiar longitudinal mass, which
is critical for understanding the unusual electron transport in graphene.

\section*{Acknowledgments}

CKU acknowledges support from the National Science Foundation under Grant No. EEC-1446038.
JS acknowledges support from the National Science Foundation under Grant Nos. CMMI-1635520 and CBET-1706815.
JS and RS acknowledge start-up funding from the Materials Science
and Engineering department at Rensselaer Polytechnic Institute.

\makeatletter{} 
\end{document}